\documentclass[fleqn,usenatbib]{mnras}

\usepackage{newtxtext,newtxmath}

\usepackage[T1]{fontenc}

\DeclareRobustCommand{\VAN}[3]{#2}
\let\VANthebibliography\thebibliography
\def\thebibliography{\DeclareRobustCommand{\VAN}[3]{##3}\VANthebibliography}

\usepackage{graphicx}	
\usepackage{amsmath}	

\title[B0823+26 variability]{Study of the variability and components of the pulsar B0823+26 at a frequency of 111 MHz}

\author[B0823+26 variability]{
M.O. Toropov$^{1}$,
S.A. Tyul'bashev$^{2}$\thanks{E-mail: serg@prao.ru (SAT)},
T.V. Smirnova$^{2}$,
V.A. Samodurov$^{3,2}$,
and I.L.Ovchinnikov$^{4}$
\\
$^{1}$ LLC TEK Inform, Moscow, 117246, Russia\\
$^{2}$ Lebedev Physical Institute, Astro Space Center, Pushchino Radio Astronomy Observatory, Pushchino, Moscow reg., 142290, Russia\\
$^{3}$ National Research University Higher School of Economics, Moscow, 109028, Russia\\
$^{4}$ Skobeltsyn Institute of Nuclear Physics, Lomonosov Moscow State University,
Moscow, 119234, Russia\\
}

\date{Accepted XXX. Received YYY; in original form ZZZ}

\pubyear{}

\begin{document}
\label{firstpage}
\pagerange{\pageref{firstpage}--\pageref{lastpage}}
\maketitle

\begin{abstract}
Studies of the pulsar B0823+26 have been carried out using the Large Phased Array (LPA) radio telescope. At time span of 5.5 years, the amplitudes of the main pulse (MP), postcursor (PC) and interpulse (IP) were evaluated in daily sessions lasting 3.7 minutes. It is shown that the ratio of the average amplitudes of MP in the bright (B) and quiet (Q) modes is 60. For B-mode, the average ratio of MP amplitudes to IP amplitudes is 65, and the ratio of MP amplitudes to PC amplitudes is 28. The number of sessions with a nulling is 4\% of the total number of sessions. Structure function (SF) and correlation function analysis of MP, IP and PC amplitude variations of over a long-time interval allowed us to detect typical time scales $37\pm 5$ days and one year. The analysis of time variations shows that the time scale of 37 days is well explained by refraction on inhomogeneities of interstellar plasma, which is distributed mostly quasi-uniformly in the line-of-sight. This scintillation makes the main contribution to the observed variability. Analysis of the structure function showed that there may be a few days variability. This time scale does not have an unambiguous interpretation but is apparently associated with the refraction of radio waves on the interstellar medium. One-year variability time scale has not been previously detected. We associate its appearance with the presence of a scattering layer on a closely located screen at a distance of about 50-100 pc from the Earth.

\end{abstract}




\section{Introduction}



Over the past 10 years, a number of papers have been published on the study of individual pulses of PSR B0823+26 (J0826+2637) (\citeauthor{Young2012}, \citeyear{Young2012}; \citeauthor{Sobey2015}, \citeyear{Sobey2015}; \citeauthor{Hermsen2018}, \citeyear{Hermsen2018}; \citeauthor{Basu2019}, \citeyear{Basu2019}; \citeauthor{Rankin2020}, \citeyear{Rankin2020}). 
Observations of B0823+26 at a frequency of 1,400 MHz (\citeauthor{Young2012}, \citeyear{Young2012}) showed that there are short (minutes) and long (hours) nulling. The rapid change between the states "radio emission is on" or "nulling is on" occurs in one rotation of the pulsar. No changes were found in the rotation speed when the pulsar is in the active or passive radiation stage. In observations with LOFAR (\citeauthor{Sobey2015}, \citeyear{Sobey2015}) a quiet B0823+26 pulsar Q-mode was discovered. Nulling was observed in both Q-mode and B-mode, but the duration of nulling in the Q-mode was 40 times longer than in the B-mode. On average, the values of the signal to noise ratio (S/N) in the two modes differ by 2 orders of magnitude (\citeauthor{Sobey2015}, \citeyear{Sobey2015}). Previously, the Q-mode was considered as nulling, since due to the weakness of radiation in this mode, it was not possible to distinguish the Q-mode from nulling.  As in the paper \citeauthor{Sobey2015} (\citeyear{Sobey2015}), observations by \citeauthor{Basu2019} (\citeyear{Basu2019}) at the frequency 320 MHz had shown, that nulling in B-mode is short and accounts for several percent of the total duration of the mode, and in Q-mode it reaches 90 percent (\citeauthor{Basu2019}, \citeyear{Basu2019}). The existence of a weak interpulse (IP) and a postcursor (PC) in the absence of the main pulse (MP) in the B-mode was also found. Simultaneous observations on radio telescopes LOFAR, Westerbork, Lovell, Effelsberg at frequencies from 135 MHz to 2.7 GHz showed that the transition between the Q-mode and B-mode occurs synchronously during one pulsar period. The observed integral flux density of B0823+26, according to \citeauthor{Bilous2016} (\citeyear{Bilous2016}), reaches 1~Jy in the meter wavelength range, which makes this pulsar one of the strongest pulsars in the northern hemisphere.


In addition to studies of pulsar pulsed radiation itself, a number of studies have been directed to the study of slow pulsar flux variations (\citeauthor{Kaspi1992}, \citeyear{Kaspi1992}; \citeauthor{Gupta1993}, \citeyear{Gupta1993}; \citeauthor{Smirnova1998}, \citeyear{Smirnova1998}; \citeauthor{Stinebring2000}, \citeyear{Stinebring2000}; \citeauthor{Daszuta2013}, \citeyear{Daszuta2013}). These variations are associated with the passage of radiation through inhomogeneities of interstellar plasma and, as was shown in \citet{Rickett1984}, they are caused by refractive scintillation on large-scale inhomogeneities. Refractive scintillation modulate diffraction scales, leading to their change over time (\citeauthor{Kaspi1992}, \citeyear{Kaspi1992}). Monitoring of 9 nearby pulsars was carried out by \citeauthor{Gupta1993} (\citeyear{Gupta1993}) at a frequency of 74 MHz for 400 days. Pulsar B0823+26 was on this list and the modulation index and refractive scintillation time were measured for it, although with large errors: $m_{\textrm{ref}} = 0.16 \pm 0.05$ and $T_{\textrm{ref}} = 12 \pm 6$ days.
Two series of observations were carried out by \citeauthor{Bhat1999} (\citeyear{Bhat1999}) at 327 MHz for pulsar B0823+26. 
The refractive scintillation time, $T_{\textrm{ref}}$,  3.6 and 3.7 days and the modulation index: $m_{\textrm{ref}} = 0.36$ and $0.32$ were measured in these series respectively. The value of $m_{\textrm{ref}}$ is in a reasonable agreement with Kolmogorov spectrum prediction model ($m_{\textrm{ref}} = 0.35$) of an extended medium.

The purpose of this paper was to study the pulsar B0823+26, both on long-term (days, weeks, months) and short-term (no more than a few minutes) observation intervals using data obtained in the monitoring program that began in 2014.

\section{Observations and initial data processing}
\label{sec2}

Using the Large Phased Array (LPA) radio telescope of the Lebedev Physical Institute of the Russian Academy of Sciences (LPI) sky survey aimed at searching for pulsars and transients (\citeauthor{Tyulbashev2016}, \citeyear{Tyulbashev2016}; \citeauthor{Tyulbashev2018}, \citeyear{Tyulbashev2018}) is being conducted. Pulsar B0823+26 is in the survey area (Pushchino Multibeams Pulsar Search - PUMPS), in which monitoring (daily) observations are carried out at an interval of more than five years. The central frequency of observations is 110.3~MHz, the receiver band $B$ = 2.5~MHz. It is divided into 32 frequency channels with a width of 78 kHz, the sampling time is 12.5 ms. The amplification in the frequency channels may differ, therefore, to compensate for this difference, a signal of a known temperature (calibration step) is used, which is recorded 6 times a day and passes through all antenna paths, starting with amplifiers at the output of dipole lines (\citeauthor{Tyulbashev2020}, \citeyear{Tyulbashev2020}).

Data are written on the hard disk for all frequency channels in hourly segments. The accuracy of the time mark of the first recording point is determined by the atomic frequency standard, but during an hour period, the time of sampling the next point is determined by a quartz oscillator, the accuracy of which is approximately $\pm$~25ms per hour interval. Due to the low accuracy of the time marking of the data, we can not do timing and sum up data for several days of observations to increase sensitivity, since the pulse phase is lost. However, the duration of pulsar recording in one session is 3.7 minutes (422 pulses) at half the power of the antenna pattern and the phase shift of the signal during this time is insignificant.

\begin{figure}
	\includegraphics[width=0.9\columnwidth] {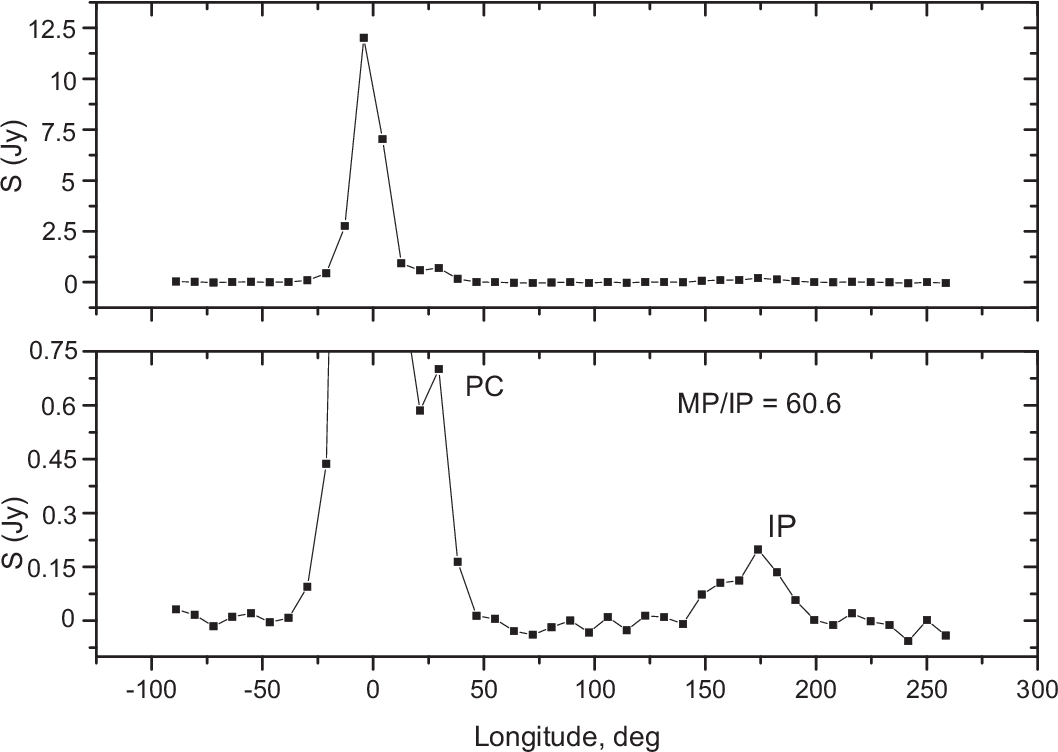}
    \caption{The upper panel shows the average profile PSR B0823+26, MJD = 57,765. The lower panel shows the same profile, increased in scale by 16 times. PC and IP are marked. The longitude in degrees is given on the $x$-axis, and the amplitude in Jy is given on the $y$-axis.}
    \label{fig:fig1}
\end{figure}

Daily observations of the pulsar B0823+26 in monitoring mode have been going on since August 2014. 
In this paper, we use observations from August 2014 until December 2019 (MJD = 56,890 - 58,847). According to the ATNF catalogue\footnote{https://www.atnf.csiro.au/research/pulsar/psrcat/} (\citeauthor{Manchester2005}, \citeyear{Manchester2005}), the pulsar period is $P_0$=0.5306 s, the dispersion measure DM=19.4 pc/cm$^3$. The dispersion smearing in the band of one channel is 9.4 ms, that is, less than the sampling time of the point.   The average pulsar profile covers 42 points per the pulsar period. The sampling time is comparable to the width of the average profile at the level of $1/2$ of the maximum ($W_{0.5}$ = 9 ms at our frequency) and therefore any details concerning the characteristics of individual pulses disappear, however, PC and IP in the average profiles per session can be seen and analyzed. The diffractive scintillation does not affect the pulse amplitude variations, since the frequency scale of scintillation $f_{\textrm{dif}}$ = (1 - 1.4) kHz at the close frequency 103 MHz (\citeauthor{Smirnova1992}, \citeyear{Smirnova1992}) is significantly less than the receiver bandwidth.

Our antenna receives linearly polarized radiation, so it is necessary to evaluate the effect of polarization on the change in the amplitude of the average profile from session to session. A  rotation measure (RM) for B0823+26 is RM = 5.9 rad/m$^2$. In the paper (\citeauthor{Sobey2015}, \citeyear{Sobey2015}) the value of the linear polarization degree $P_\textrm{L} = 15 \pm 5\%$ is obtained for the average profile at the frequency 143 MHz. The period of the Faraday rotation of the polarization plane at the frequency $f$ = 110~MHz ($B \ll f$) is: $P_\textrm{F} [\textrm{kHz}] = \pi f^3 \times 10^5/(18 \textrm{RM}$[rad/m$^2$]), $P_\textrm{F} = 3.94$~MHz. Here $f$ is observation frequency in $10^2$ MHz. This period exceeds the receiver band, and with a degree of polarization of about $15\%$, polarization has little effect on the variations from session to session in the amplitude of the average pulsar profile. The contribution of the ionosphere to modulation of the flux density does not exceed $10\%$ and can be neglected.

The processing of observation data consisted of a series of sequential steps: calibration was carried out using a calibration step; the noise level was estimated and if it was high, the data was not processed; the baseline was subtracted, the gain in the channels was equalized, dispersion compensation was carried out; all pulses were recorded on a disk and then the average profile was determined by summing them up with a given period. In the resulting average profile, and circular shift was carried out so that MP was at a distance of $1/4$ phase of the period from the beginning of the average profile, in this case IP is at a distance of $3/4$ phase of the period; S/N was calculated for MP, PC and IP. According to strong sources observed in two adjacent beams, the state of the ionosphere was monitored, and if the coordinate shift in declination was large, then the processed records were not used for the analysis. 11.7\% of days  were removed due to poor ionospheric conditions and
interference.  Twice the antenna was under technical verification and did not work on observation: JD = 57589 -57608, JD = 57611 - 57631.

As a result of processing of all observation sessions, files were created in which arrays of individual pulses for each observation session were placed, as well as baseline values and root-mean-square (rms) deviations ($\sigma_\textrm{n}$), determined for each pulse outside MP and IP longitudes. The average profile obtained by summing up all the individual pulses was recorded in the same file. The drawings of the average profile for all frequency channels and the summed-up profile were formed, allowing to assess the quality of observations. Based on the results of processing for all available days that were not distorted by interferences and the ionosphere, additional tables were created containing the values of MP, PC and IP amplitudes and the values $\sigma_\textrm{n}$ for each day.

As an example, Fig.\ref{fig:fig1} shows a typical average profile of the pulsar B0823+26 after processing observations according to the scheme presented above. The profile shows MP, PC and IP. The zero longitude corresponds to the position of MP center. The amplitudes MP, PC and IP defined as the values of the maxima in the corresponding narrow specified ranges of longitudes of the average profile. The chosen ranges of longitudes from $-30$ to $+23$ deg. for MP, from $+21$ to $+38$ deg. for PC and from $+149$ to $+191$ deg. for IP.  The width of MP at the level of $1/2$ of the amplitude is equal to: $W_{0.5}$ = 19 ms, IP is 56 ms. The ratio of their widths $r$ = 3, which corresponds to the measurements in the paper \citeauthor{Basu2019} (\citeyear{Basu2019}) at the frequency 320 MHz: that is $r$ = 3.3. Naturally, the obtained pulse widths are significantly larger than their actual widths due to the smearing in the receiver band and coarse time sampling, but for us it does not matter. Though dispersive smearing is comparable to time sampling it acts equally on all components of the profile decreasing its amplitudes but it doesn't affect their amplitude variations in time and also amplitudes ratio. The non-optimal observation mode does not affect relative amplitudes of MP, PC and IP.

\section{Analysis and results}

\subsection{Intrinsic emission modes}

\begin{figure}
	\includegraphics[width=0.9\columnwidth]{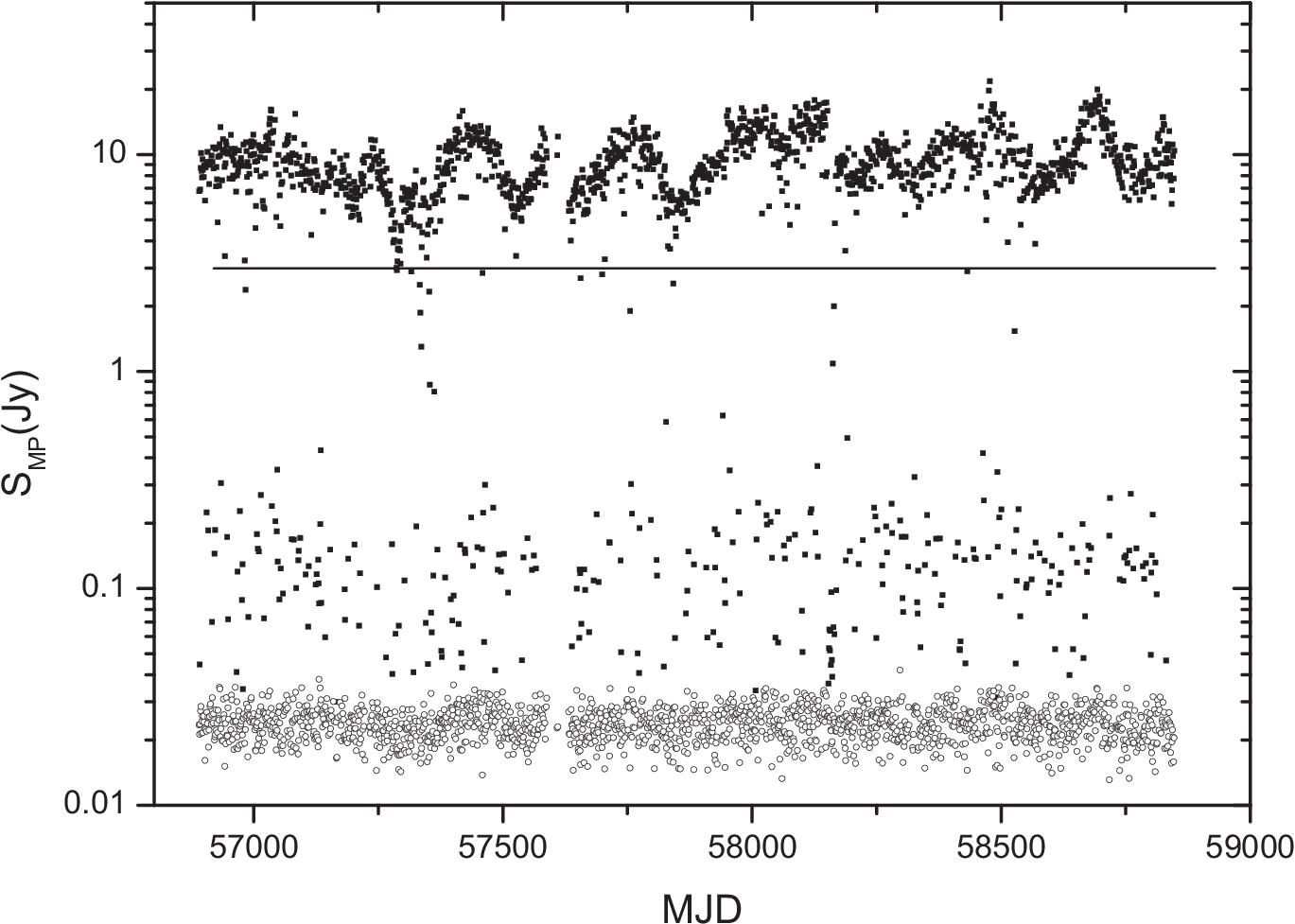}	
    \caption{Amplitude variations in Jy of MP vs. time (21.08.2014 - 30.12.2019). Axis $y$ is amplitude in log scale, axis $x$ is time in Julian days (MJD). The lower part of the figure (hollow circles) shows the change in $\sigma_\textrm{n}$ during observations. The straight line corresponds to the level of $S_{\textrm{MP}} = 3$~Jy.}
    \label{fig:fig2}
\end{figure}

Fig.\ref{fig:fig2} shows the change in the amplitude of MP in Jy from session to session over the entire observation period and $\sigma_\textrm{n}$ (lower part of the figure). 
Shown on the Fig.\ref{fig:fig2} $\sigma_\textrm{n}$ is the average value obtained from all 422 individual pulses in each observation session. The average value of $\sigma_\textrm{n}$ over the entire observation interval: $\langle {\sigma_\textrm{n}}\rangle = 25 \pm  4$~mJy. The main part of the MP amplitudes has $S_{\textrm{MP}}$ values $>$ 3~Jy and has significant variations over time. This is a rather sharp boundary, within which $83\%$ of all data is located. The other highlighted part has an upper boundary $A_{\textrm{MP}} < 0.5$~Jy and this includes sessions in which a calm Q-mode and nulling are included. A small part of the data (0.5~Jy$<S_{\textrm{MP}}<$3~Jy, see Fig.\ref{fig:fig2}) in the intermediate domain, corresponds, apparently, to cases when both modes are realized. The sessions with $S_{\textrm{MP}} > 3$~Jy correspond to the flash B-mode. For Q-mode, we assume that 0.5~Jy$> S_{\textrm{MP}} > 3\langle {\sigma_\textrm{n}}\rangle$ (0.075~Jy). Then the number of sessions in Q-mode, $N_\textrm{Q}$, is equal to 218 out of 1,692, i.e. $12.9\%$. The ratio of the average values of the amplitudes in B and Q modes  $\langle{A_\textrm{B}}\rangle/\langle{A_\textrm{Q}}\rangle$ is 60. We will assume that the days when $S_{\textrm{MP}}$ is less than 3$\langle{\sigma_\textrm{n}}\rangle$ correspond to nulling. The ratio of pulses with $S_{\textrm{MP}} < 3\langle{\sigma_\textrm{n}}\rangle$ to the total number of sessions),  $N_{\textrm{nul}}$, is $4\%$, and for amplitudes less than 4$\langle{\sigma_\textrm{n}}\rangle$ it is  $5.5\%$. The ratio of the pulse amplitudes are $\langle{S_{\textrm{MP}}\rangle}/\langle{S_{\textrm{PC}}\rangle}=28.4$ and $\langle{S_{\textrm{MP}}\rangle}/\langle{S_{\textrm{IP}}\rangle}=65$. In the paper \citeauthor{Rathnasree1995} (\citeyear{Rathnasree1995}) according to observations at frequencies of 430 and 1,400 MHz, it was shown that the nulling of this pulsar occupy $6.4\pm 0.8\%$ of the entire recording. The determination of the number of nulling depends on the sensitivity of the antenna and the determination of the upper boundary for the amplitude of the noise. 

\begin{figure}
	\includegraphics[width=0.9\columnwidth]{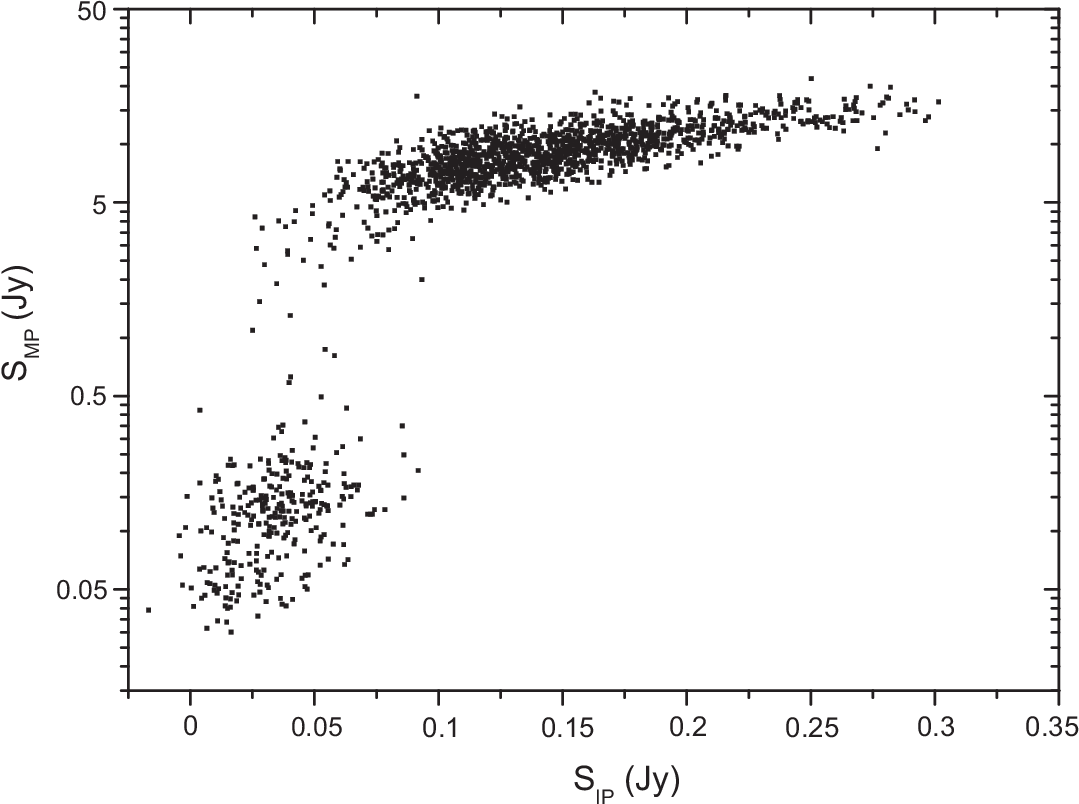}
    \caption{The relationship between MP and IP amplitudes in Jy. Logarithmic scale is used for $y$ axis.}
    \label{fig:fig3}
\end{figure}

\begin{figure}
	\includegraphics[width=0.9\columnwidth]{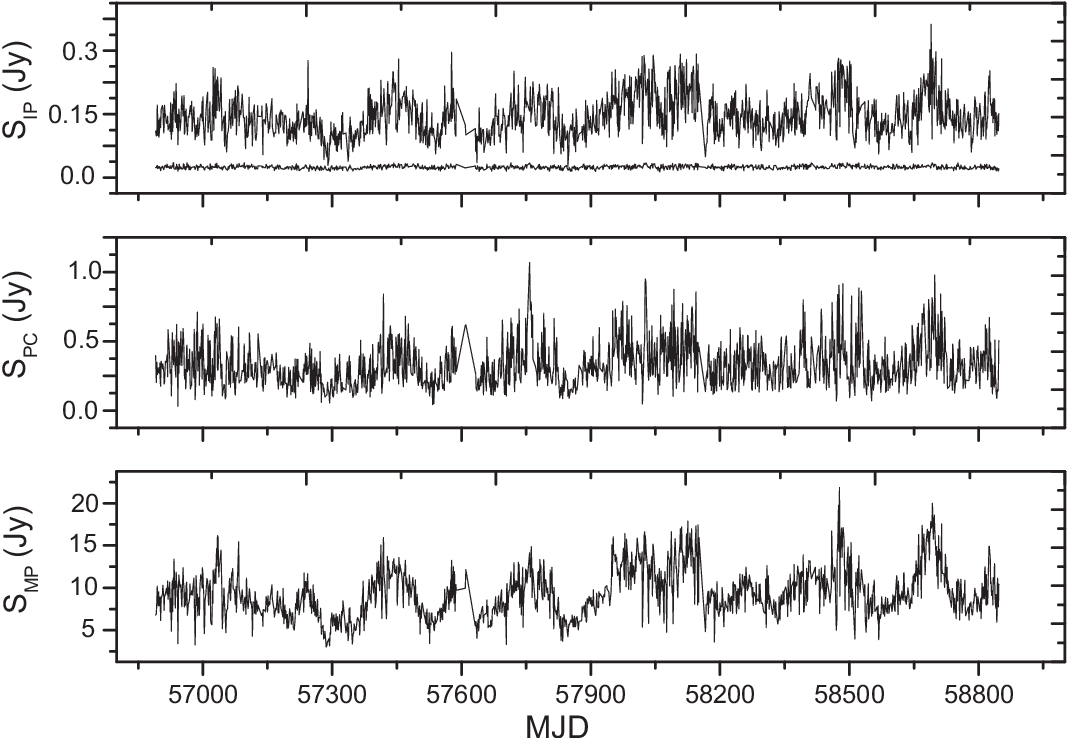}
    \caption{Amplitude variations MP, PC and IP with a time in Jy. Only sessions in which $S_{\textrm{MP}} >$~3~Jy  (B-mode) are selected here. The upper part of the figure also shows the change of $\sigma_\textrm{n}$ vs. time.}
    \label{fig:fig4}
\end{figure}

Fig.\ref{fig:fig3} shows the relation between the amplitudes of MP and IP. For this figure, values of maximum amplitudes at MP and IP longitudes for average profiles are used, for which at MP longitude the amplitude is higher than $\langle{\sigma_\textrm{n}}\rangle = 25$~mJy. Here you can also see the division of points into 2 groups. The first group is a cloud of dots in the lower left corner of the figure. It is related to the Q-mode  ($S_{\textrm{MP}} < 0.5$~Jy) and, accordingly, we see MP, but we do not see IP at the level above 3$\langle{\sigma_\textrm{n}}\rangle$. The formal values of $S_{\textrm{IP}}$ with $S_{\textrm{IP}}< 3\langle {\sigma_\textrm{n}}\rangle$ are just noise. The second group of points is located at the top part of Fig.\ref{fig:fig3}, and for $S_{\textrm{MP}} > 3$~Jy, there is such a dependence, which is more clearly seen in Fig.\ref{fig:fig4}. Since we do not see interpulses in Q-mode, all further analysis is carried out for the data in B-mode.  Fig.\ref{fig:fig4} shows the variations in the amplitudes of MP, PC and IP vs. time. 

Obviously, the variations of all three components are correlated. The correlated amplitude variations are caused by the passage of radiation from the pulsar through the inhomogeneities of the interstellar plasma and are not related to the intrinsic variability of the pulsar. These are refractive scintillations on large plasma inhomogeneities, which will be considered later.


The average values of the ratio of the amplitudes of the average profile components obtained from these data are: $\langle{S_{\textrm{MP}}}/{S_{\textrm{IP}}\rangle} = 66 \pm 15$, $\langle{S_{\textrm{MP}}/S_{\textrm{PC}}\rangle} = 33 \pm 15$ and they are in good agreement with the ratios of average values given above.

\begin{figure}
	\includegraphics[width=1.0\columnwidth]{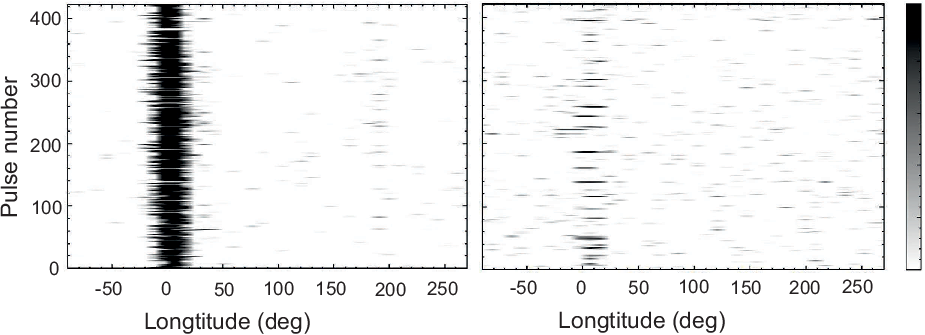}
    \caption{The left panel shows radiation of B0823+26 in B-mode (the session MJD 58,116). The right panel shows the pulsar radiation in Q-mode (the session MJD 57,757). The pulse number is shown on the $y$-axis, longitude in degrees is shown on the $x$-axis. The color corresponds to the range of amplitudes of individual pulses from 3$\sigma_\textrm{n}$ (white color) up to the maximum value 7$\sigma_\textrm{n}$ (black color).}
    \label{fig:fig5}
\end{figure}

Fig.\ref{fig:fig5} shows the characteristic variations of the individual pulses intensity in B and Q modes for two sessions. The number of pulses with no emission in B-mode ($S < 3\sigma_\textrm{n}$, nulling) is usually less than  $5\%$. In Q-mode, nulling prevails, thus is usually more than  $85\%$ of time.

\begin{figure}
	\includegraphics[width=0.9\columnwidth]{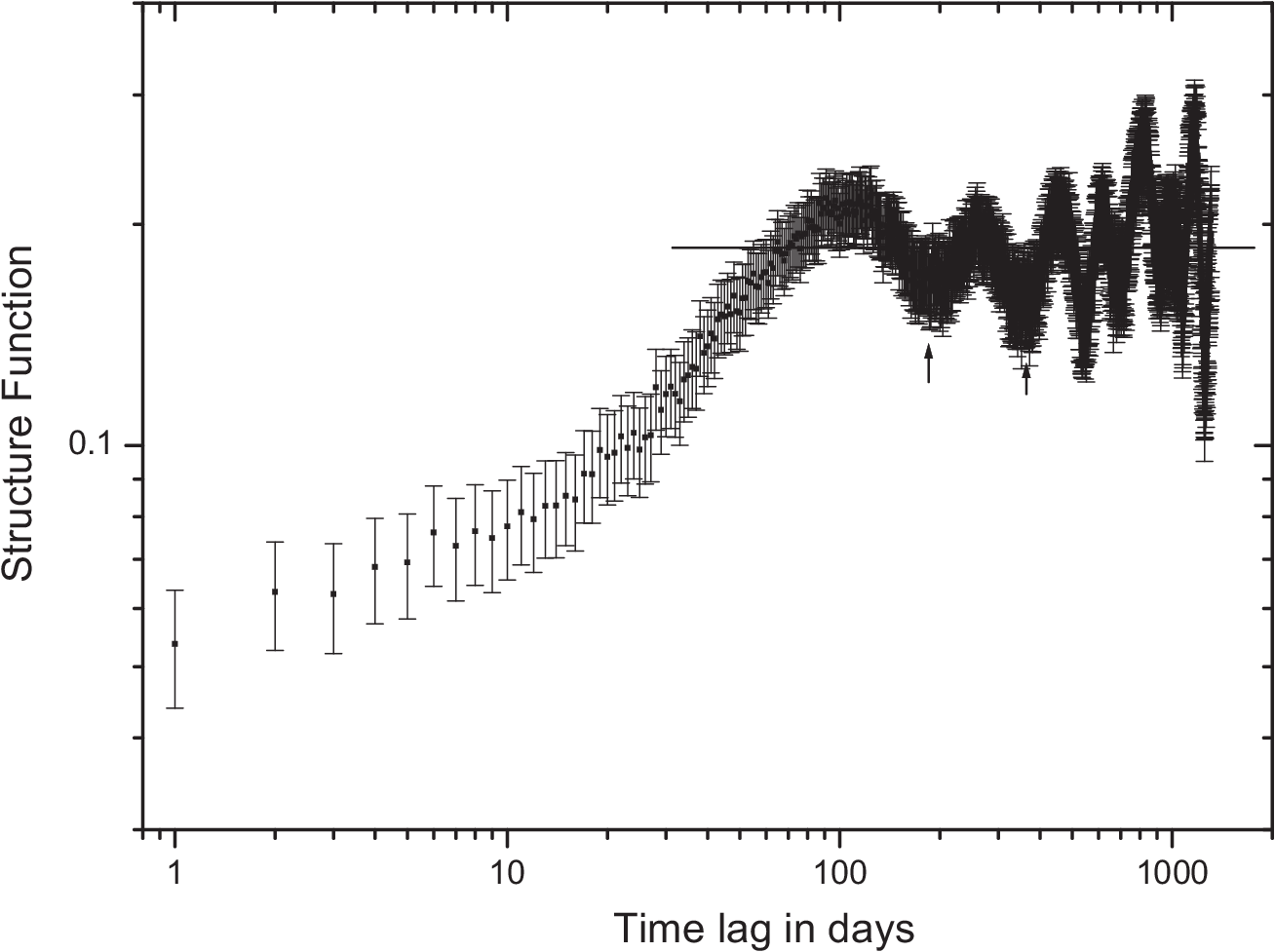}
    \caption{SF of amplitude modulation for PSR B0823+26 is shown on a double logarithmic scale. The $x$-axis is the time lag in days, $y$-axis is the value of SF. The arrows indicate lags of 180 and 360 days. Stright line shows the level of 2$m^2$}.
    \label{fig:fig6}
\end{figure}

\begin{figure*}
	\includegraphics[width=0.8\textwidth]{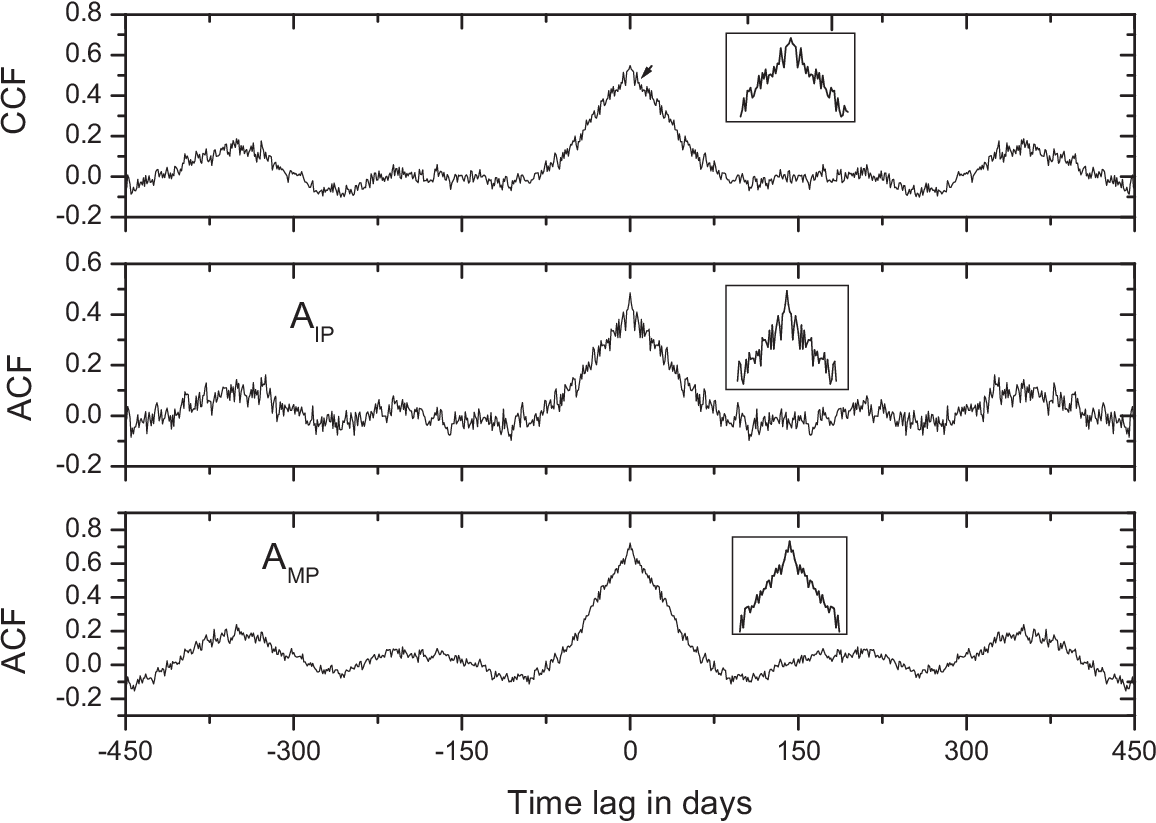}
    \caption{Normalized ACF for MP amplitude variations (lower panel) and IP amplitude variations (middle panel). On the top panel, there is CCF between the amplitudes of MP and IP. The $x$-axis shows the time lag in days.}
    \label{fig:fig7}
\end{figure*}

\subsection{Extrinsic flux density variations}

Radio waves are scattered into an angular spectrum with a width $\theta_{\textrm{scat}}$, called the scattering angle. At a certain distance $r$ from the scattering layer, mutual interference of various components of the angular spectrum leads to radiation amplitude modulation. There are diffractive scintillation on the scales $s_{\textrm{dif}} = \lambda/\theta_{\textrm{scat}}$ ($s_{\textrm{dif}} \sim 10^7$ cm per $\lambda = 1$~m) and refractive scintillation, which are realized on significantly larger spatial scales: $s_{\textrm{ref}} = \theta_{\textrm{scat}} r$ ($s_{\textrm{ref}} \sim 10^{12}$~cm). Characteristic time scales of diffractive scintillation, $t_{\textrm{dif}} = s_{\textrm{dif}}/{\bf |V_{\textrm{ef}}|}$, at the frequency 110 MHz are seconds and minutes, and for refractive scintillation: $T_{\textrm{ref}} = s_{\textrm{ref}}/{\bf |V_{\textrm{ef}}|}$ are months and years. Here ${\bf |V_{\textrm{ef}}|}$ is the velocity of the diffraction pattern movement relative to the observer. This velocity is determined as a geometrical sum of three components perpendicular to the line of sight: 

\begin{equation}
  {\bf |V_{\textrm{ef}}|} = \frac{r}{R-r}{\bf V}_{\textrm{psr}} + {\bf V}_{\textrm{obs}} - \frac{R}{R-r}{\bf V}_{\textrm{scr}},
    \label{eq:1}
\end{equation}
where ${\bf V}_{\textrm{psr}}$, ${\bf V}_{\textrm{obs}}$ and ${\bf V}_{\textrm{scr}}$ are velocities of a pulsar, the Earth and the screen, correspondingly, $R$ is the distance from the observer to the pulsar, $r$ the distance from the observer to the screen. The distance to the pulsar $R$ is 500 pc, the tangential velocity of the pulsar is 272.3 km/s (\citeauthor{Deller2019}, \citeyear{Deller2019}). Both modes of scintillations are realized on meter wavelengths.

The time of diffractive scintillation at a frequency of 103 MHz for PSR B0823+26 is: $t_{\textrm{dif}} = 35$s (\citeauthor{Smirnova1992}, \citeyear{Smirnova1992}), and these variations are smoothed out both by the receiver band and by averaging the pulses when obtaining an averaged profile. This time scale is  in good agreement with multiple measurements reported in \citeauthor{Bhat1999} (\citeyear{Bhat1999}), which would scale down to $\sim 35-40$ seconds at $\sim 110$~MHz. The frequency scale of diffractive scintillation, as noted Sec.~\ref{sec2}, is significantly less than the width of the frequency band (2.5 MHz). Accordingly, the diffractive scintillation will be smoothed out by the receiver band and will not affect the amplitude variations of the averaged profile. If the scattering layer is very close to the Earth ($r \ll R$), then the velocity ${\bf V}_{ef}$  is determined by the velocity of the Earth, and if the scattering occurs close to the pulsar, then it is determined by the velocity of the pulsar. The velocity of the screen is about 10 km/s (\citeauthor{Linsky2008}, \citeyear{Linsky2008}), and the velocity of the Earth is about 30 km/s. The case of an extended medium corresponds to $r = R$ and ${\bf V}_{\textrm{ef}} = {\bf V}_{\textrm{psr}}$.

As can be seen from Fig.\ref{fig:fig4}, the time scale of variations is tens of days. Therefore, we believe that this is refractive scintillation. 

\subsubsection{Structure function}
To estimate the time scale of amplitude variations, a structure function (SF) is usually used. It is defined as follows (\citeauthor{Stinebring2000}, \citeyear{Stinebring2000}): 

\begin{equation}
SF(k)= \frac{1}{\langle{S}\rangle ^2 N(k)} \sum_{i=1}^{M-k} g(i)g(i+k)[(S(i)-S(i+k)]^2,
    \label{eq:2}
\end{equation}
where $\langle S \rangle$ this is the average value of the amplitude, $M$ is the length of the array, $k$ is the time shift in days, $k = 1, 2, ..., 0.8M$, $g(i) = 1$, when there is a value for a given day $i$, and it is equal to 0 otherwise. $N(k) = \sum g(i)g(i+k)$. SF defined by all data for which $S_{MP} > 3$~Jy is shown on Fig.\ref{fig:fig6}.  The time scale is usually determined by the time lag at which it falls by a factor of 2. According to the definition of the structure function (equation~\ref{eq:2}), for large time shifts the amplitude variations should be independent in the absence of correlated variations at these shifts and reach a constant level of $2m^2$ defined by modulation index $m$. 
The modulation index is defined as:

\begin{equation}
m = \sum_{i=1}^N \left[\frac{(S-\langle{S}\rangle )^2}{N-1}\right]^{1/2}/\langle{S}\rangle.
    \label{eq:3}
\end{equation}

The average value of MP amplitude determined over the entire observation range is: $\langle S_{\textrm{MP}}\rangle = 9.4\pm 2.85$~Jy. Here the error corresponds to the rms deviation and is determined mainly by strong modulation associated with refractive scintillation. The mean integral flux density at 111 MHz is $1.3\pm 0.4$~Jy which is comparable with \citeauthor{Bilous2016} (\citeyear{Bilous2016}) (1 Jy). The modulation index determined by variations in the amplitude of MP for all days (for sessions with $S_{\textrm{MP}} > 3$~Jy) is equal to $m = 0.3\pm 0.06$. Here the error is determined by the rms deviation for 5 independent intervals of 300 days. In Fig.\ref{fig:fig6}, the straight line shows the level corresponding to the value of $2m^2$ = 0.18 determined by us. This level corresponds well to the constant respect to which variations of the structure function take place. The local minima with scales of 180 and 360 days relative to this level are shown by arrows. SF has the bending on the time shift of about 10-15 days and 2 linear sections that correspond to two time scales. A shorter scale of several days has a significantly smaller amplitude. Uncertainties in the structure function shown in Fig.\ref{fig:fig6} were calculated as $\Delta SF(k) = \sigma_n[8SF(k)/N(k)]^{0.5}$ (\citeauthor{Simonetti1985}, \citeyear{Simonetti1985}). Additional data analysis was performed to see bending in SF. We divided our time interval into 3 adjacent sections of 400 days each and calculated SF for each one. For all independent samples there is a change in the slope of SF on a scale of the order of 15 days. Since the modulation index has a large error, then determining the lag at which SF drops by a factor of 2 will also have a large error. It is better to use autocorrelation and cross-corellation functions (ACF; CCF) to define refractive time scale. In this case it is needed just to measure the width of ACF. Below we will define $T_{\textrm{ref}}$ from the correlation function. Local minima are visible on SF in the area of large time lags, corresponding to lags of 180 and 360 days. They are caused by quasi-periodic amplitude modulation with such characteristic periods.

\subsubsection{Correlation analysis}
To determine the characteristic time scales of amplitude variations, we also used correlation analysis. Fig.\ref{fig:fig7} shows normalized ACF from MP and IP amplitude variations as well as CCF between MP and IP amplitudes. For ACF, the value for zero lag is removed. On all three panels of Fig.\ref{fig:fig7} you can see the characteristic semi-annual and annual time scales of pulse amplitude variations  as well as a narrow detail of small amplitude. If the variations with different time scales are independent, then their relative contribution can be estimated. The sum of the squares of their amplitudes at zero time lag is equal to the value of CCF. We have determined the ratio of the variation amplitudes by 2.7 times, and the scales themselves as: the short one 5 days about and the long one of $T_{\textrm{ref}} = 37 \pm 5$ days. The short scale was determined as a time lag corresponding to 1/2 of the lag for bending of the correlation function shown by the arrow on Fig.\ref{fig:fig7}. Although the scale of several days has a significantly smaller relative contribution to the amplitude variations, it manifests itself in all three correlation functions (Fig.~\ref{fig:fig7}) and therefore we think that it is possibly exists. We estimated the statistical error in the correlation functions as 8\% for the  time shift of 5 days. In fact, the error may be greater, since the level of the CCF bending is accurate to several days. Therefore, we can only talk about a short scale of a few days.

Figure 4 in \citeauthor{Gupta1993} (\citeyear{Gupta1993}) presents the structure function of PSR B0823+26 at 74 MHz. It shows 2 scales, one of which is defined as $T_{\textrm{ref}} = 12\pm 6$ days. Converteted to a frequency of 111~MHz in accordance with the dependence $T_{\textrm{ref}} \sim f^{-2.2}$  for Kolmogorov spectrum it will be 5 days which is the same as we have. The long scale was determined by the drop of CCF in 2 times from the level at the bend visible on the scale about 5 days. The relative error in $T_{\textrm{ref}}$ we have determined as proportional to $(T_{\textrm{ref}}/T)^{0.5}$, where $T$ is total observation interval.  Let us note that when smoothing MP amplitude variations by 5 points (Fig.\ref{fig:fig4}), the modulation index becomes equal to $m = 0.25$ and the value determined from the SF is: $T_{\textrm{ref}} = 40$ days, which coincides with the value of CCF.

\section{Discussion of the results}

Our study of B0823+26 showed that it has B, Q modes and nulling. Table \ref{tab:table2} shows fractions of time occupied by these modes and nulling in observations on LPA at the frequency 111 MHz, on LOFAR at the frequency 143 MHz (\citeauthor{Sobey2015}, \citeyear{Sobey2015}), on GMRT at the frequency 320 MHz (\citeauthor{Basu2019}, \citeyear{Basu2019}). The total fraction of nulling and observation sessions when there is the transition from B-mode to Q-mode took place was $4.1\%$, which together with the fraction of B and Q-mode brings the value of the total number of sessions to $100\%$.

\begin{table}
	\centering
	\caption{Fractions of time in B/Q modes and fraction of nulling in these modes.}
	\label{tab:table2}
	\begin{tabular}{cccc} 
		\hline
	  &LPA  & LOFAR & GMRT \\
	\hline	
B-mode      & 83\%    & 63-90\%   & $>$77\% \\
Q-mode      & 12.9\%  & 10-37\%   & $<$23\% \\
nulling (B) & 4\%  & $\ge$ 1.8\% & 3.4-4.4\% \\
nulling (Q) & 88\%   & $\ge$ 80\%  & $>$91\%   \\
		\hline
	\end{tabular}
\end{table}

As can be seen from Table~\ref{tab:table2}, in general, the fraction of time occupied by the B and Q modes, as well as the fraction of nulling in these modes, are comparable at all three frequencies. The ratios of peak flux densities in B and Q modes at close frequencies of 111 MHz and 143 MHz differ considerably. In our observations, this ratio is 60 times, in LOFAR observations it is 100 times. However, this ratio is conditional and the division into B and Q modes depends on the amplitude limit we have adopted (less than 0.5 Jy) for cases of Q-mode. If we make this level lower, then, accordingly, the ratio of amplitudes and the fraction of time spent in the B-mode will be higher. The average values of the relative amplitude of IP and PC turned out to be in our study: $1.5 \pm 0.3\%$ and $3 \pm 1.3\%$. At 143 MHz, the corresponding values for IP are $0.9\pm 0.05\%$ and $3.76\pm 0.05\%$ for PC. 

Refractive and diffractive scintillations of the pulsar B0823+26 were observed at a frequency of 1.7 GHz (\citeauthor{Daszuta2013}, \citeyear{Daszuta2013}). During the observation interval from 2003 to 2006, 70 sessions were conducted with a duration of sessions from 6 to 12 hours. It was shown that the time scales for diffractive scintillation are 19.3 minutes, and for refractive scintillation 144 minutes. The time of refractive scintillation depends on the frequency of observation, as: $T_{\textrm{ref}} \sim f^{-\alpha/(\alpha-2)}$, where $\alpha$ is the power-law index of the spatial spectrum of interstellar plasma inhomogeneities. In the area between the inner and outer scales: $s_{\textrm{inn}} < s < s_{\textrm{out}}$, the spectrum is described by a power law (\citeauthor{Rickett1990}, \citeyear{Rickett1990}): $P_{3n(\rho)} = C_n ^2\rho^{-\alpha}$, where $C_n ^2$ is a level of turbulence along the line-of-sight, and $\rho = 2\pi/s$. For the Kolmogorov spectrum $\alpha = 11/3$. Analysis of 2D correlation functions of the dynamic spectrum for B0823+26 gives the value $\alpha = 3.66$ (\citeauthor{Popov2021}, \citeyear{Popov2021}), which is close to the Kolmogorov value of $3.67$. We will use the value $\alpha = 3.67$, then $T_{\textrm{ref}} \sim f^{-2.2}$. Recalculation of 144 minutes at a frequency of 1,700 MHz to our frequency of 111 MHz gives 40.5 days, which coincides with the value we have obtained ($37\pm 5$ days). \citeauthor{Bhat1999} (\citeyear{Bhat1999}) determined the refractive scale for the pulsar B0823+26 at a frequency of 327 MHz as $T_{\textrm{ref}} = 3.6$ and $3.7$ days for two series of observations. When converted to our frequency, it is 39 and 40 days, which also corresponds well to our value. For comparison, we put the measured values of $T_{ref}$ at different frequencies and their extrapolated values ($T_{ref,ext}$) to  frequency of 111 MHz in Table~\ref{tab:table3}. The content of columns is the following: (1) - frequency of observation, (2) - measured refractive time, (3) - reference, (4) - extrapolated to f = 111MHz refractive time.

\begin{table}
	\centering
	\caption{Estimated and extrapolated refractive time}
	\label{tab:table3}
	\begin{tabular}{cccc} 
		\hline
$f$, MHz      & $T_{ref}$   & Reference & $T_{ref,ext}$ \\
		\hline
74 & $12\pm 6$~days & \citeauthor{Gupta1993} (\citeyear{Gupta1993}) & $5\pm 2.5$~days \\
111 & few days; $37\pm 5$~days& this paper& \\
327 & 3.6~days & \citeauthor{Bhat1999} (\citeyear{Bhat1999}) & 39 days \\
1700& $144\pm 23$~min & \citeauthor{Daszuta2013} (\citeyear{Daszuta2013}) & $40.5\pm 6.5$~days \\
		\hline
	\end{tabular}
\end{table}

In the paper \citeauthor{Fadeev2018} (\citeyear{Fadeev2018}), from observations on a ground-space interferometer at a frequency of 324 MHz, the scattering angle in the direction of B0823+26: $\theta_{\textrm{scat}} = 0.77\pm 0.08$ mas was determined. Assuming $\theta_{\textrm{scat}}\sim f^{-2.2}$, we will obtain on our frequency $\theta_{scat} = 8.13$ mas. On the other hand, $\theta_{\textrm{scat}}$ for an extended medium model and Kolmogorov spectrum can be estimated from the relation $\theta_{\textrm{scat}} = (c/\pi Rf_{\textrm{dif}})^{1/2}$ (\citeauthor{Bhat1999}, \citeyear{Bhat1999}). At our frequency $f_{\textrm{dif}} = 2$ kHz, and accordingly  $\theta_{\textrm{scat}} = 9.7$ mas. The characteristic scale of refractive scintillation: $T_{\textrm{ref}} = r \theta_{\textrm{scat}}/{\bf |V_{\textrm{ef}}|}$. Using the resulting value $T_{\textrm{ref}} = 37$ days and $\theta_{\textrm{scat}} = 9.7$ mas, we will determine the relation: $r/{\bf |V_{\textrm{ef}}|}  = T_{\textrm{ref}}/\theta_{\textrm{scat}} = 6.7\times 10^{13}$~s. For the case of an extended scattering medium, assuming that the pulsar velocity significantly exceeds ${\bf |V_{\textrm{obs}}|}$ and ${\bf |V_{\textrm{scr}}|}$ , for $r = R = 500$ pc, ${\bf |V_{\textrm{ef}}|} = {\bf |V_{\textrm{psr}}|} = 272.3$ km/s, we will get:  $r/{\bf |V_{\textrm{ef}}|} = 5.5\times 10^{13}$~s which coincides well with the above relation. Consequently, it can be concluded that the extended medium makes a significant contribution to scattering. In the paper \citeauthor{Fadeev2018} (\citeyear{Fadeev2018}) the distance to the scattering screen was estimated from the analysis of the secondary spectrum as $r = 240$ pc. The arcs in the secondary spectrum are strongly blurred, which may serve as evidence that the thin screen does not make the main contribution to scattering.

\citeauthor{Romani1986} (\citeyear{Romani1986}) found that complex scattering geometries such as one or more thick screens or an extended medium will give rise to larger flux modulations than those caused by a thin-screen model. According to the authors, if the scattering is uniformly distributed along the line of sight, the flux fluctuations will be larger by a factor of 2.3 compared to the thin-screen model, in the case of a Kolmogorov spectrum. In the work of \citeauthor{Bhat1999} (\citeyear{Bhat1999}), the modulation index of refractive scintillation was measured for PSR B0823+26 in two series of observations, as 0.36 and 0.32. These values agree well with the predicted value for the extended medium and the Kolmogorov spectrum: $m_{\textrm{ref}} = 0.35$. For the thin screen model, the predicted value is 0.13. For the extended medium model and the Kolmogorov spectrum, the modulation index is determined by the relation \citeauthor{Rickett1984} (\citeyear{Rickett1984}):
\begin{equation}
m_{\textrm{ref}} = 1.21(f_{\textrm{dif}}/2f)^{0.17}.
    \label{eq:4}
\end{equation}

At a frequency of 111 MHz, we obtained the value $m_{\textrm{ref}} = 0.25$ after smoothing the amplitude variations with a scale of 5 days and $m_{\textrm{ref}} = 0.3$ without smoothing. The expected value of $m_{\textrm{ref}}$  for this model is $m_{\textrm{exp}} = 0.17$ for $f_{\textrm{dif}} = 2$~kHz. The value of the modulation index obtained by us agrees quite well with the assumption that the extended medium makes a significant contribution to scattering. The obtained values for $m_{\textrm{ref}}$ at frequencies of 327~MHz \citeauthor{Bhat1999} (\citeyear{Bhat1999}) and \citeauthor{Daszuta2013} (\citeyear{Daszuta2013}) at 1700~MHz also supported this. Perhaps several screens located on the line of sight are equivalent to the action of an extended medium.

The time scale of about 5 days, which we found in SF and correlation functions (ACF, CCF), has a small contribution (about $1/3$) to the variations in the amplitude of pulsar radiation and may be associated with refractive scintillation on a structure whose spatial scale is about 7 times smaller than the main scale. The structures (plasma inhomogeneities) that cause scintillation on the five-day scale can be explained by an additional thin screen on the line-of-sight. The contribution of refraction of this structure to the observed overall picture of variability is small. \citeauthor{Gupta1993} (\citeyear{Gupta1993}) observations at 74~MHz also show a short scale of flux variations comparable to ours when recalculated to 111~MHz. However, the S/N ratio was not high in these observations for B0823+26. For an unambiguous interpretation, observations at other low frequencies with a good S/N are needed.

The semi-annual and annual time scales observed in SF, ACF and CCF, indicate their probable connection with the Earth's orbital motion. Perhaps there is a scattering layer in the direction of the pulsar at a close distance from the Earth, and as the line-of-sight crosses different points of the orbit, ${\bf |V_{\textrm{ef}}|}$ changes in accordance with the equation~\ref{eq:1}. In order to have a noticeable influence of this effect, it is necessary that the contribution from the pulsar velocity into ${\bf |V_{\textrm{ef}}|}$: $r{\bf |V_{\textrm{psr}}|}/(R-r)$ is comparable to the Earth velocity. Accordingly, the distance to such a layer should be about 50-100~pc. Then the contribution of the pulsar velocity will be 30-68~km/s. Annual variations of the amplitude in the direction of B0823+26 were not previously noted by other authors.

\section{Conclusion}
Daily observations at the LPA LPI over a long-time interval allowed us to confirm the existence of a Q-mode for the B0823+26 pulsar at the frequency of 111 MHz. As in the observations of other authors (\citeauthor{Sobey2015}, \citeyear{Sobey2015}; \citeauthor{Basu2019}, \citeyear{Basu2019}) we note that the pulsar has B and Q modes, as well as nulling in both modes. The duration of nulling in Q-mode is many times longer than in B-mode.

Analysis of the peak amplitude variations of the average pulse over a long-time span: from August 2014 to December 2019, allowed us to detect 3 time scales associated with the passage of radiation through the inhomogeneities of interstellar plasma.

\section*{Acknowledgements}
The study was carried out at the expense of a grant Russian Science Foundation 22-12-00236, https://rscf.ru/project/22-12-00236/.

\section*{Data availability}
The PUMPS survey not finished yet. The raw data underlying this paper will be shared on reasonable request to the corresponding author. The table with amplitudes of MP, PC, IP and rms in http://prao.ru/online\%20data/onlinedata.html

\end{document}